# Exceptional broadband absorption of nanoporous gold explained by plasmonic resonances at dangling ligaments

**MUHAMMAD SALMAN WAHIDI,**[1,*] **MAURICE PFEIFFER,**[1] **XINYAN WU,**[1] **FATEMEH EBRAHIMI,**[1,2] **MANFRED EICH,**[1,2] **AND ALEXANDER YU. PETROV,**[1,2]

[1]*Institute of Optical and Electronic Materials, Hamburg University of Technology, Germany*
[2]*Institute of Functional Materials for Sustainability, Helmholtz-Zentrum Geesthacht, Germany*
*[\*muhammad.wahidi@tuhh.de](*muhammad.wahidi@tuhh.de)*

**Abstract:** Nanoporous gold (npAu) has emerged as a potential candidate for many optical applications exploiting its large surface to volume ratio and high broadband absorption. However, the physical origin of its enhanced visible and near infrared absorption remained unclear and till now was not explicable by simplified models. Here, we have employed leveled-wave approximants to simulate the optical response of realistic npAu structures. First, our simulations reproduced well the experimental absorption spectra. Second, we identify multiple resonances in the gaps between dangling ligaments that occur at the top and bottom surfaces of npAu films as the key contribution to the broadband absorption. These resonances at the surface of npAu cannot be captured by bulk effective medium models and should be considered separately as a surface effect. The additional absorption due to dangling ligaments contributes up to 70 % to overall absorption of npAu. Our results provide deeper insights into the absorption behavior of npAu, indicating promising avenues for photocatalysis and sensing applications.

## I.     Introduction

NpAu consists of a continuous network of gold ligaments and interconnecting pores that can be tuned from approximately 5-6 nanometer up to micrometer size by controlled dealloying and thermal coarsening.[1–4] The npAu thin films have high surface to volume ratio, excellent conductivity, stability, and biocompatibility which make it appealing for various applications in sensing, drug delivery, and electrochemical applications.[4–8] In the optical regime, npAu with small ligament size appears black owing to its broadband absorption over visible and near infrared (NIR) wavelength ranges.[9–14] In addition, optical properties of npAu were tailored electrochemically or by atomic layer deposition coatings.[15–18] The plasmonic properties of npAu have been effectively exploited for surface-enhanced Raman scattering (SERS), which indicates strong field concentration.[19–21] The nanostructure of npAu has been shown to enhance catalytic activity by light irradiation.[22–26]

Though the structure of npAu and its experimental optical response are relatively well known there is still no consensus on the effective dielectric properties of npAu that would explain the measured transmission and reflection of npAu thin films. One approach is to fit the npAu as a homogeneous medium with Drude and Lorentz terms.[27–32] These phenomenological models rely on several fitting parameters, such as effective plasma frequency, and have been able to find good fits to the reflectance and transmittance measurements from various npAu samples. However, they fail to capture the connection between the structural and optical properties of npAu and thus to predict optical properties from the known structural parameters. Especially the applied Lorentz terms help to fit the observed spectra but are not attributed to resonances at particular structural features of npAu. Thus, the physical origin of the high broadband absorption of npAu is not explained by these models.

A more physical explanation of the observed optical properties would be given by effective medium models. These models rely on the averaging of field parameters over the representative structure and can predict expected optical properties from known geometrical parameters such as filling fraction, average pore or ligament geometry, and others. The models might differ depending on the averaging approach and geometry approximations. Two effective medium models with different averaging approach, Bruggeman and Maxwell-Garnett (MG), are widely used for this purpose. In the Bruggeman formulation each of the materials is considered as an inclusion in the average medium itself.[33] As the Bruggeman model assumes a homogenous mixture of two materials, both inclusions, air and gold in npAu, are typically assumed to be spherical.[11,12,34,35] This assumption contradicts the geometrical properties of npAu, which is a connected network of gold ligaments and not a collection of isolated gold spheres, and thus the Bruggeman model cannot give a correct physical picture. An attempt to use anisotropic inclusions in Bruggeman model considers the distinct transmission spectra of npAu films as originating from two localized surface plasmon resonances (LSPRs) orthogonal and parallel to the ligaments, which are approximated by nanorods with fitted aspect ratio.[36] Though the resonance orthogonal to the ligaments has a clear physical origin, the assumption that connected ligament structure can be approximated by nanorods with termination at their ends is questionable. Also, the Bruggeman approach of putting gold inclusions into the average medium seems not to be a very good approximation for npAu structure as ligaments of npAu are mostly surrounded by pores and not the average medium. The MG model on the other hand allows a clear choice of what material is inserted and considers the other material as its matrix.[33] The MG model was applied to npAu films with small filling fraction of gold ~30 %.[37–39] In Jalas et al.,[37] the MG model was implemented by considering ligaments as cylindrical wires. This model splits the Au volume fraction into randomly oriented infinite wires, which are to $1/3^{rd}$ parallel and $2/3^{rd}$ orthogonal to the incident electric field (Supporting Information Note 1). The resonances orthogonal to the ligaments, which are also seen in the experimental curves of npAu as distinct dip in transmission at approx. 500 nm, are accurately reproduced by this model. Also, this model predicts well the reflection in the long wavelength range due to assumption of a connected geometry. However, it failed to explain the strong absorption of npAu in the intermediate regime from 500 to 1000 nm wavelength.

In this study we draw attention to the top and bottom surfaces of npAu films where dangling ligaments exist that can result in LSPR in the gap between two neighboring ligament ends. These resonances were previously observed on npAu particles.[40–42] In particular Vidal et al., demonstrated that the optical properties of Au nanosponges are strongly influenced by spatially and spectrally LSPRs.[43,44] However, such resonances were previously observed on isolated npAu particles and were not seen in experiments with npAu films due to averaging over large area. It was also not clear to which extent these resonances contribute to the absorption properties of npAu thin films. As shown previously,[32] the LSPRs at the surface of npAu were explicitly excluded considering npAu films sandwiched between dielectric substrates and showing that effective parameters are not changing. In recent works, local cathodoluminescence spectroscopy measurements of npAu film might also indicate existence of local resonances in npAu.[45,46]

In this study we used synthetic leveled-wave structures[47–50] to emulate npAu geometry and simulated representative periodic volumes to study LSPR appearance. We observed prominent resonances within the visible and NIR wavelength regime. The origin of these pronounced resonances from individual samples was clearly attributed to the gaps between dangling ligaments at the top and bottom surfaces of the npAu films. Computational constraints necessitate the use of small sample volumes in simulations. To bridge the gap and achieve spectra comparable to experimental observations from large npAu areas, we employed ensemble averaging across multiple samples. Further, we present a comparative analysis of surface resonance contributions using the MG model developed by Jalas et al. as a baseline.[37] Notably, the contribution of surface resonances leads to significantly higher absorption in the visible and NIR regions, highlighting their critical role in explaining optical properties of npAu thin films.

## II. Structure generation

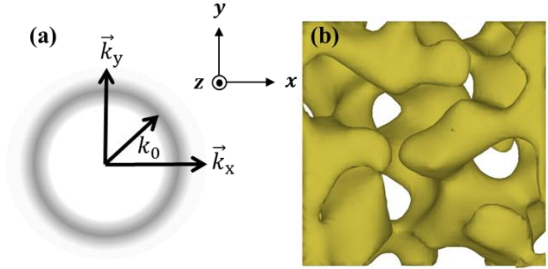

Figure 1. (a) Schematic of a 2D cross section of a spherical shell in the reciprocal space with radius $k_0$ and center at the origin. The shell has a Gaussian intensity profile along radial direction. (b) A 3D figure of generated nanoporous gold (npAu) approximant, which is periodic along $x$ and $y$ direction, whereas the film normal is along $z$ −axis, which is also the excitation direction.

Nanoporous gold structures were computationally generated from reciprocal space, employing a leveled-wave approach.[47–49,51] The approach is based on defining a distribution $P(k)$ in reciprocal space that resembles the structure factor of npAu, assigning a random phase factor to each point in reciprocal space and converting it into real space by inverse Fourier transform. The obtained Gaussian random field in real space is binarized by defining a cutting level, where the gold structure takes the volume of the random field above the level, thus the name leveled-wave structure.[47,49,50] To limit the range of density correlations we take a continuous distribution of wave numbers. Such a limit on range of density correlations is also evident from small angle X-ray scattering (SAXS) investigations showing a wide ring from npAu structure in reciprocal space.[3,52] We have modelled it as a spherical shell in reciprocal space with a radius $k_0$ and the shell defined by a Gaussian function with standard deviation $\sigma_k$. We estimated $\sigma_k$ by fitting the peak width of the SAXS data ($\sigma_k$ determination is discussed in the Supporting Information Note 2).[3] To numerically generate structures the reciprocal space is discretized, and each diametrically opposite point pair is assigned with a random phase $\phi$. The real part of the inverse Fourier transform results in the random field function in the real space with a Gaussian distribution of the amplitude. To obtain a bicontinuous structure, the random field is binarized using a cutting level $\xi$ defined by the required volume fraction of gold:[47]

$$\xi = \sqrt{2}\sigma_A \cdot \text{erf}^{-1}(1 - 2f_{Au}) \qquad (1)$$

Where $f_{Au}$ is the volume fraction of Au and $\sigma_A$ is the standard deviation of the Gaussian field amplitude in real space. In this study, we required periodic structures in two directions to emulate a continuous film by periodic boundary conditions. The periodicity is implemented by using a fast Fourier transform on a grid in reciprocal space that has discretization $dk = 2\pi/L_s$, where $L_s$ is the side length of the periodic volume.

Characteristic period or wavelength is another crucial metric of npAu structure, which is defined as $a = 2\pi/k_0$.[50] The characteristic wavelength $a$ can be used to derive the characteristic length $\tilde{L} = 1.23a$, which quantifies average distance between ligament centers.[50] The mean ligament diameter $L$ depends on the characteristic length and filling fraction. We have determined it for our experimental samples applying the AQUAMI software to scanning electron microscopy (SEM) images of npAu.[53] It generates a distance map over the whole image area and returns average ligament diameter of $L = 0.57a$, which is larger than reported.[50] Note that they only considered the average ligament neck diameter. In simulated structures, we set the $f_{Au}$ to ~25% and $k_0 = 0.1256$ nm$^{-1}$ which results in $a = 50$ nm and $L \simeq 28.5$ nm.

The npAu cubes prepared via the described method exhibit well-defined, sharp periodic boundaries in all directions. However, for numerical models of npAu thin films, we introduce a correction at the top and bottom boundaries, where periodic boundary condition is not applied. The smoothening of sharp edges is achieved by multiplication of the Gaussian random field with a window function which gradually goes to zero at the top and bottom surfaces. This step introduces a controlled level of rounding of the dangling ligaments bringing the numerical model closer to the observed characteristics of npAu surface (see Supporting Information Note 3).

### III.    Numerical simulations

After the generation of the npAu approximants, an STL file was generated bound to the original 3D grid. Meshlab, an open access software,[54] was employed to obtain smooth surfaces and reduce the size of the STL files employing marching cube algorithm.[55] This significantly decreased the number of mesh cells in simulations and hence simulation time without altering the topology of these approximants. The frequency domain (FD) solver of commercially available CST Studio Suite 2025 was then used to simulate the optical properties.[56] The FD solver uses finite element method, which allows tetragonal conforming meshing of the npAu structure. This is an important advantage over the finite difference time domain (FDTD) method and its orthogonal mesh, that might lead to electric field artifacts at the discretized boundary between dielectric and metal.[57] Plane wave excitation along $z$ direction was used (see Figure 1b). Periodic boundary conditions were implemented on the lateral boundaries of the periodic npAu approximant. The simulated npAu approximant is free standing in air with refractive index $n = 1$, and Au properties were taken from Ref.[58]

In Figure 2a-c, the absorption spectra of three unique realizations with $L_s = 200$ nm and thickness $h = 150$ nm are shown. The films are modelled by first defining a film of initial thickness of 200 nm from which a film thickness of 160 nm is achieved by cutting. Then, this geometry is multiplied by the window function along z-direction with transition regions equal to the half of the average ligament diameter $L/2$ at the top and bottom surfaces (see Supporting Information Note 3). This results in the final thickness of ~150 nm, thus reduced the initial thickness of 160 nm by approx. 5 nm on each side by removing the sharp edges at the two respective boundaries. Each unique configuration is obtained by a new random phase distribution in reciprocal space. Each of the samples shows distinct resonances in the visible and NIR wavelength range (absorption spectra of more samples can be found in Supporting Information Note 4). The absorption spectrum calculated from the effective medium parameters predicted by the MG model[37] and corresponding film thickness acts as a baseline, whereas on top of this baseline the contribution of resonances can be evaluated. Similar resonances were observed in the scattering spectra of npAu particles[42–44] and cathodoluminescence spectra of npAu films[45] and particles.[46]

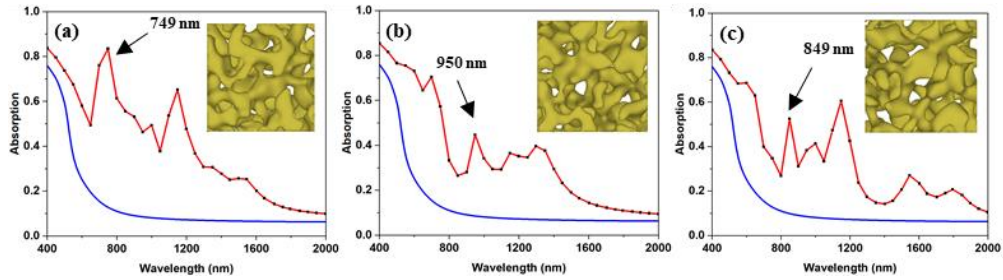

Figure 2. (a-c) Three unique npAu approximants, although statistically equivalent, show distinct absorption spectra (red with black dots indicating the simulated discrete wavelengths points) across the visible and NIR wavelength range with clear resonances. Each sample's absorption spectra are compared to the absorption spectra predicted by the Maxwell-Garnett effective medium model (blue), which serves as a reference and, by definition, does not show resonances originating from the film surfaces.

Upon analysis of the electric field distributions at resonance wavelengths, we have confirmed that these resonances originated at distinguished gaps between two dangling ligaments at the top and bottom surfaces of the npAu approximants. Figure 3a-c shows the distribution of the electric field amplitude on the gold surface at three resonant wavelengths 749 nm, 950 nm and 849 nm of the three samples shown in Figure 2, correspondingly. Plane wave excitation along $z$ direction was used with a linearly $y$-polarized electric field. Spatially localized electric field enhancement can be observed in all three cases at the tips of two dangling ligaments opposite to each other (more examples are presented in the Supporting Information Note 5). We also compare electric field distributions at resonant and close-by wavelength to confirm that field enhancements are corresponding to resonances. In a study by Maaroof et al,[32] the LSPRs at the surface on npAu were explicitly excluded considering npAu films sandwiched between dielectric substrates and showing that effective parameters are not changing. However, the observed LSPRs are spatially localized between the dangling ligaments (as shown in Figure 3a-c) with electric fields located mostly inside the film. Even if the substrate would slightly shift the resonance frequencies, this will not cause the resonances to vanish and, on average, will not alter the number of resonances or their strength.

To understand the appearance of the resonances and explain their spectral position, we consider similar resonances observed in the gap between two plasmonic structures, where shape of the structures and the gap size govern the optical response.[59,60] We see, for example, that the resonance shifts to longer wavelength with decreasing gap size. This is also demonstrated by simulating a grid of gold cylinders with a varying gap (see in Supporting Information Note 6). From Ref.[59] it can be seen that such gap resonance can be maximally shifted to 800 nm wavelength in antenna of 200 nm length. It seems the npAu structure provides a much larger size of the effective antenna and can demonstrate resonances even above 1600 nm as can be seen in Figure 2c and Supporting Information Note 6. Alternatively, the gap between two dangling ligaments can be seen as a part of the vertical split-ring resonator embedded into the npAu. Such resonators are also known for their long wavelength resonances.[61,62]

Due to the inherent variability and limited size of numerically generated npAu structures, the individual npAu realizations in simulations exhibit distinct absorption spectra. To approach the experimental spectra which are collected from square centimeter areas, we averaged the absorption spectra of 15 npAu samples with $200 \times 200 \times 150$ nm$^3$ volume. The averaged absorption is shown in Figure 4. It is evident that npAu numerical model shows higher absorption than predicted by the effective medium model at wavelengths above 500 nm due to the resonances at the surface in the visible and NIR wavelength range.

We should note that resonances are sometimes difficult to identify in the spectra as they might overlap or be strongly broadened. To identify the contribution of resonances at a particular wavelength, we have analyzed the average depth profile of absorption for 15 samples at 700 nm wavelength. As shown in Figure 5, there is an extraordinary absorption concentration at the top 45 nm layer of the samples. The additional absorption in this 45 nm layer above the average absorption in the bulk of the samples corresponds to approximately 25% of total absorption. This is not so much as the difference in absorption between simulation and effective medium model observed in the spectra of Figure 4—at 700 nm. It means that LSPRs at the surface of npAu not only absorb locally but also redirect incident radiation into the film and thus lead to additional absorption in the bulk. This coupling between LSPRs at the surface and bulk npAu film should be further investigated.

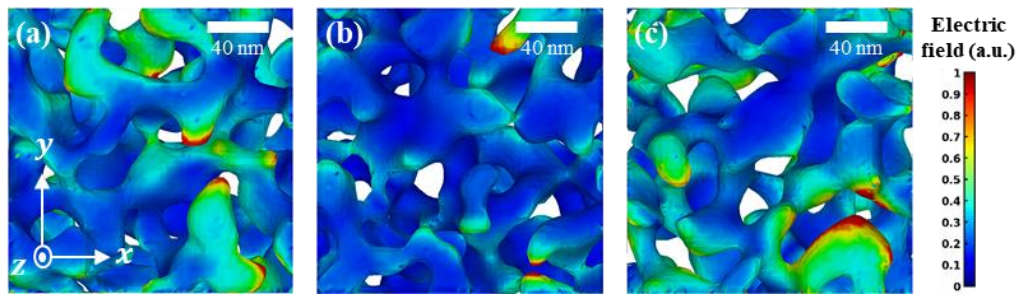

Figure 3. (a-c). Simulated electric field distributions for three distinct npAu approximants corresponding to absorption peaks at wavelength (a) 749 nm (b) 950 nm (c) 849. Here, a projection of the view of the top surface into the film is shown with the k-vector of the excitation wave normal to the plane of the diagram and pointing into the structure. The electric field is polarized along $y-$ axis. The amplitude of the electric field in the metal at the air-gold interface is shown, to underline the field enhancement in the 3D structure. NpAu samples are shown at resonant wavelengths for the samples shown in Figure 2. (a-c).

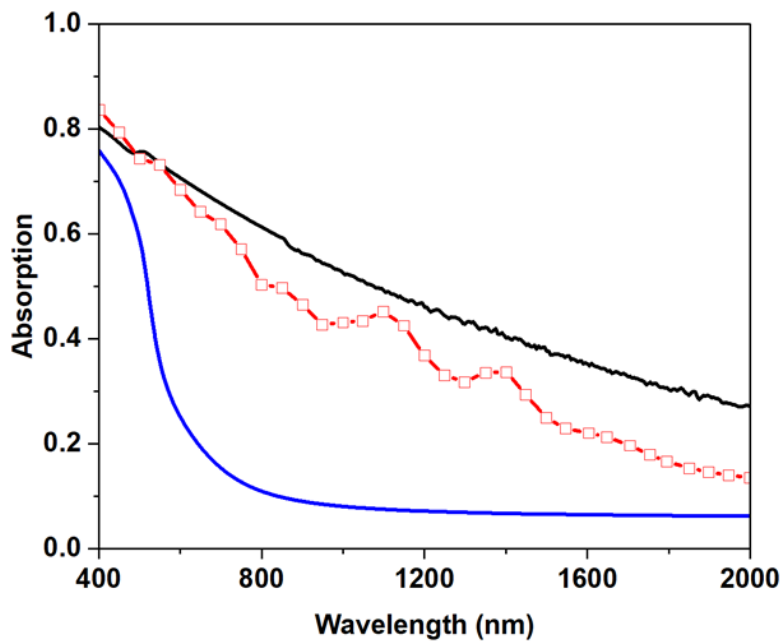

Figure 4. The averaged simulated absorption spectra (red with white boxes indicating the simulated discrete wavelengths points) of 15 distinct npAu realizations. Experimental (black line) absorption spectra of npAu films (~150 nm thickness) with ~25% Au filling fraction. These aggregated absorption spectra are compared to the absorption predicted by Maxwell Garnett model (blue), which serves as a reference.

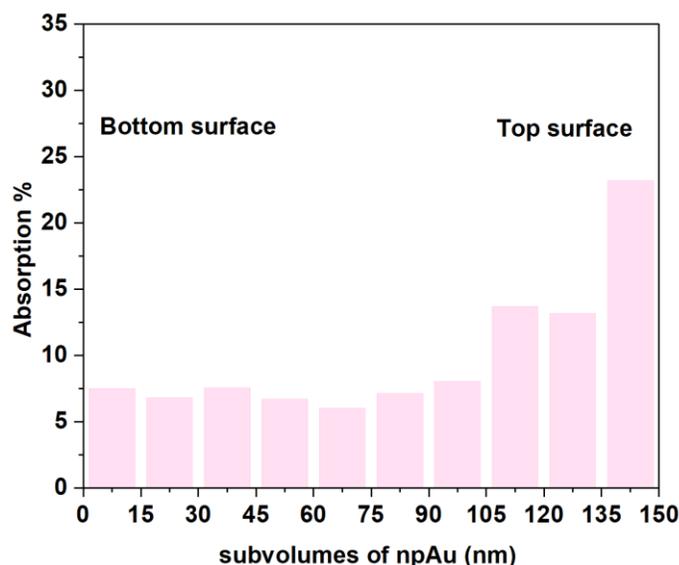

Figure 5. Histogram of the absorption distribution versus depth for 15 npAu samples at a wavelength of 700 nm discretized in 15 nm layers and normalized to 100%.

## IV.   Comparison to experiment

NpAu thin films were fabricated through electrochemical dealloying of 6 −carat white gold leaf precursors (Gerstendoerfer, Germany) at room temperature (21 ℃). The composition was verified by energy-dispersive X-ray spectroscopy (EDS), which determined the gold atomic percentage content to be approximately 18%. The dealloying process was carried out using a three-electrode setup, where the gold leaf served as the working electrode, a coiled silver wire (diameter 0.1 mm) applied as the counter electrode, and a pseudo-Ag/AgCl reference electrode (0.53 V vs. reversible hydrogen electrode (RHE) in 1 M $HClO_4$) was employed. All potentials presented in this work are reported relative to the RHE. The supporting electrolyte used was 1M $HClO_4$ (70%, p.a., Merck, Germany) prepared with ultrapure water (18.2 MΩ, Sartorius). A potentiostat (Autolab PGSTAT204, Metrohm, Netherlands) controlled the potential for all the dealloying experiments.

Dealloying was conducted using a two-step potentiostatic method. First, a potential of 1.26 V was applied until the current dropped below $2 \times 10^{-5}$ A, reducing the risk of excessive structural deformation such as shrinkage or cracking. The potential was then increased to 1.36 V to enhance silver removal. After dealloying, the film was carefully rinsed in ultrapure water then caught by a quartz glass (25.4 × 25.4 × 1 mm, Plano) for further optical measurements. The quartz glasses were prior washed in ethanol in sonic bath for 10 minutes then dried by air flow.

SEM was employed to characterize the structure of the npAu thin film, revealing a thickness of approximately 140 nm and Au solid fraction of ~25%. The optical properties of such samples were further investigated using UV-Vis spectroscopy while the sample was on a quartz substrate (Supporting Information Note 7). Due to observed inhomogeneity within the sample, diffuse transmission and reflection spectra were measured at three distinct locations using an integrating sphere of the UV-Vis spectrometer (Supporting Information Note 7). The experimental absorption spectra presented in Figure 4 were obtained by averaging the transmission and reflection spectra acquired from these three different spots.

Our numerical model exhibits good agreement with the experimentally measured spectra. It shows the long wavelength absorption that was missing in the prediction of an effective medium model. However, still a quantitative discrepancy exists between the simulated and experimental results. This discrepancy can be attributed to several effects. First, it is difficult to exactly determine the average thickness and filling fraction of npAu and there can be a systematic error of several percent. Second, the small lateral dimensions of the npAu sample in the finite element method (FEM) model might influence the statistics of resonances. In particular, long wavelength resonances might rely on the large size of an effective antenna around the gap. Additional large volume simulations are required to clarify this effect, which are currently not possible with our computational facilities. Third, the exact termination of top and bottom surface might influence the amount, geometry and size of the gaps between ligaments and thus observed number and strength of the resonances in simulations. There might be a systematic difference between leveled-wave geometry terminated by a window function that we use (see Supporting Information Note 8) and the real surface of npAu film. A rigorous experimental study of npAu geometry at the surface is required to emulate it better in a numerical model.

## V.     Conclusion

This study highlights the significant role of resonances at the film surfaces of npAu in the optical response of such npAu films. To confirm that, we used leveled-wave structural approximants of the npAu structure and simulated their absorption properties in frequency and space. These simulated samples exhibited distinct absorption spectra with prominent resonances at the visible and NIR wavelengths. By averaging the spectra across an ensemble of samples, we obtained a representative absorption profile characterized by significantly higher absorption compared to the predictions of the Maxwell-Garnett effective medium model.

Furthermore, we investigated the underlying structural features behind the observed resonances in the simulated npAu samples. Through a detailed analysis of the electric field distribution at resonant and non-resonant wavelengths, we identify the origin of these resonances as localized surface plasmon polaritons at the gaps between dangling ligaments at the npAu surface. This refined numerical model effectively captures the enhanced absorption behavior of npAu, which the effective medium model fails to account for. Further studies with larger simulation volumes and more precise description of the geometry of npAu film surface are required to obtain a quantitative fit with experiments.

The presented results also show that it is not possible to explain the optical properties of npAu by pure effective medium models that consider npAu as a medium with an effective dielectric constant. In reality the response of npAu is defined by bulk npAu and random resonances at the surface. Thus, any effective medium model that will be defined will unavoidably have thickness-dependent effective parameters, contradicting the concept of an effective medium description. We believe that a Maxwell-Garnett effective medium model developed earlier[37] still represents a reasonable approximation of bulk npAu properties. In the future a more complex analytical model for npAu films can be developed which would consider a coupled system of a homogenous effective medium film and resonances at the surface with a defined distribution and coupling constants.

**Acknowledgements.** The authors acknowledge the sponsorship from Dassault Systemes with their CST Studio Suite software and the funding from Deutscher Akademischer Austauschdienst (DAAD).

**Disclosures.**  The authors declare no conflicts of interest.

**Data availability.** The data that support the findings of this study are available from the corresponding author upon reasonable request.

**Supplemental document.** See supporting information file for further data.

# Supporting Information for Publication

# Exceptional broadband absorption of nanoporous gold explained by plasmonic resonances at dangling ligaments

MUHAMMAD SALMAN WAHIDI,[1] MAURICE PFEIFFER,[1] XINYAN WU,[1] FATEMEH EBRAHIMI,[1,2] MANFRED EICH,[1,2] AND ALEXANDER YU. PETROV,[1,2]

[1]Institute of Optical and Electronic Materials, Hamburg University of Technology, Germany
[2]Institute of Functional Materials for Sustainability, Helmholtz-Zentrum Geesthacht, Germany
*muhammad.wahidi@tuhh.de

**Supplementary Note 1: Maxwell Garnett model for nanoporous gold**

Nanoporous gold (npAu) is approximated by the effective medium model of a cubic grid of gold wires by Jalas et al.[1] This model splits the gold volume fraction into randomly oriented infinite wires, which are to 1/3$^{rd}$ parallel and 2/3$^{rd}$ orthogonal to the incident electric field. Here, $\varepsilon_{Au}$ and $\varepsilon_d$ are the wavelength-dependent permittivities of gold,[2] and air as dielectric ($\varepsilon_d = 1$), whereas $f_{Au,\perp}$, $f_{Au,\parallel}$, and $f_d$ are the volume fractions of the constituent materials. The subscripts $\parallel$, $\perp$ refer to the gold wire parallel or perpendicular, respectively. The resulting homogenized response $\varepsilon_{eff}$ corresponds to the Maxwell Garnett formulae for a cubic grid of cylindrical wires, shown in equation (1). The gold fraction is 25% and is divided into parallel and perpendicular gold wires, making up 33% and 67%, respectively.

$$\varepsilon_{\text{eff}} = \frac{f_{Au,\perp}\varepsilon_{Au}\frac{2\varepsilon_d}{\varepsilon_{Au}+\varepsilon_d} + f_{Au,\parallel}\varepsilon_{Au} + f_d\varepsilon_d}{f_{Au,\perp}\frac{2\varepsilon_d}{\varepsilon_{Au}+\varepsilon_d} + f_{Au,\parallel} + f_d} \quad (1)$$

**Supplementary Note 2: Estimation of $\sigma_k$ through SAXS data of npAu**

Small-angle X-ray scattering (SAXS) from npAu thin films yields an intensity profile $I(k)$ in reciprocal space, which we fit by a Gaussian function characterized by a peak position $k_0$ and a standard deviation $\sigma_k$. In Figure 4c of Ref.[3], time-resolved SAXS curves at multiple timestamps during dealloying are presented. To isolate the dealloying-induced contribution, the spectrum at t=0 s (parent Au–Ag alloy) is used as a baseline and subtracted from the curve at t=3300 s (fully dealloyed sample). The resulting difference spectrum is fitted with a Gaussian function (Figure S1), from which the full width at half maximum (FWHM), $\sigma_k$, and $k_0$ are extracted. The fitted $k_0$ and FWHM are reported below. $\sigma_k$ is used to generate numerical realizations as described in equation (4):

$$\text{FWHM} = 2\sqrt{2ln(2)}\sigma_k \qquad (2)$$

$$\text{FWHM} \approx (2/3)k_0 \qquad (3)$$

$$\sigma_k \approx k_0/3\sqrt{2ln(2)} \qquad (4)$$

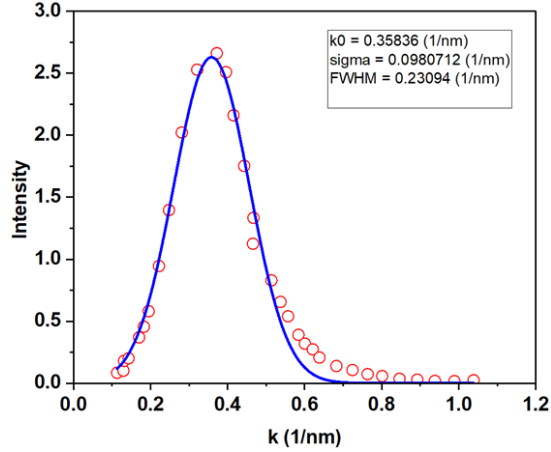

**Figure S1.** The red circles indicate the measured SAXS intensity profile of npAu (after subtraction of results from Figure 4(c) in Ref. 3) and the blue curve is the fitted Gaussian function. The inset indicates the essential parameters of this Gaussian function.

## Supplementary Note 3: Rounding off sharp boundaries in numerical modelling of npAu

As discussed in the manuscript, we numerically generate npAu through inverse Fourier transform of a spherical shell with random phase in reciprocal space. Once we transform it into real space we obtain a Gaussian random field with an amplitude $n(x, y, z)$. This function $n$ is binarized to get a bicontinuous npAu structure. If the binarized structure is terminated by a plane then ligaments at the surface have flat ends with sharp edges, as can be seen in Figure S2. To avoid flat ends, we multiply $n$ with a window function $f(z)$ which has tapers at the top and bottom surfaces of npAu film. We have set the width of taper sections as $L/2$ (half of mean ligament diameter e.g. ~14 nm here). The tapers follow a semicircle function going from 1 in the bulk npAu to 0 at the end of the taper. The window function with thickness $h_w$ is defined as:

$$f(z) = \begin{cases} \sqrt{1-(2z/L)^2}, & 0 \le z < \dfrac{L}{2} \\ 1, & \dfrac{L}{2} \le z \le h_w - \dfrac{L}{2} \\ \sqrt{1-(2(h_w-z)/L)^2}, & h_w - \dfrac{L}{2} < z \le h_w \end{cases} \qquad (5)$$

The Figure S2 illustrates the effect of rounding on sharp edges of npAu ligaments. The final thickness is $h \sim 150$ nm, reduced by approx. 5 nm on each side compared to window thickness $h_w = 160$ nm. The thickness of the film is defined as the distance between two planes where average density of npAu reduces by half of the designed value, in our case 28%. The average density is obtained by npAu integration over 2D planes orthogonal to axis $z$ (shown in Figure S3). After the generation of npAu approximants, Meshlab, an open access software,[4] is employed to reduce the size of the STL files, which further reduces the density of gold to ~25% (which is considered as the final gold fraction of the simulated npAu approximants).

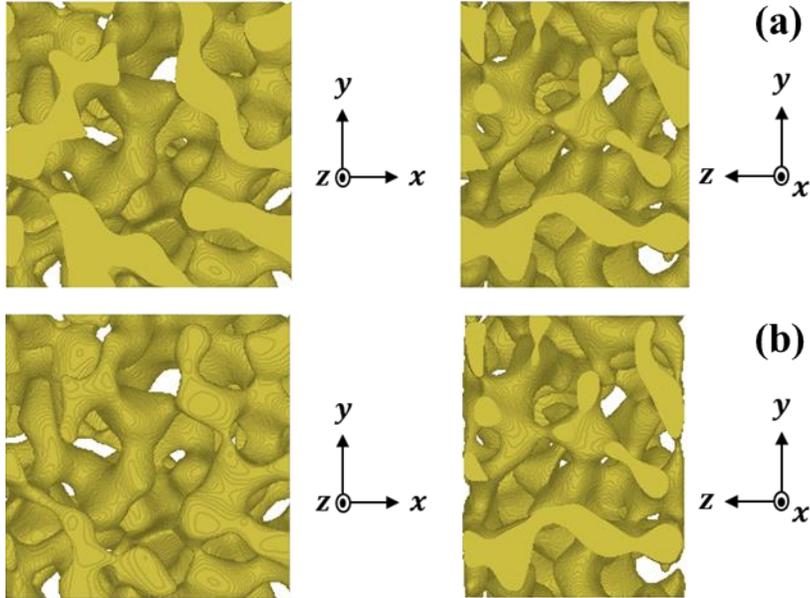

**Figure S2.** The comparison between npAu with (a) sharp boundary (top panel) and (b) rounded boundary (bottom panel)

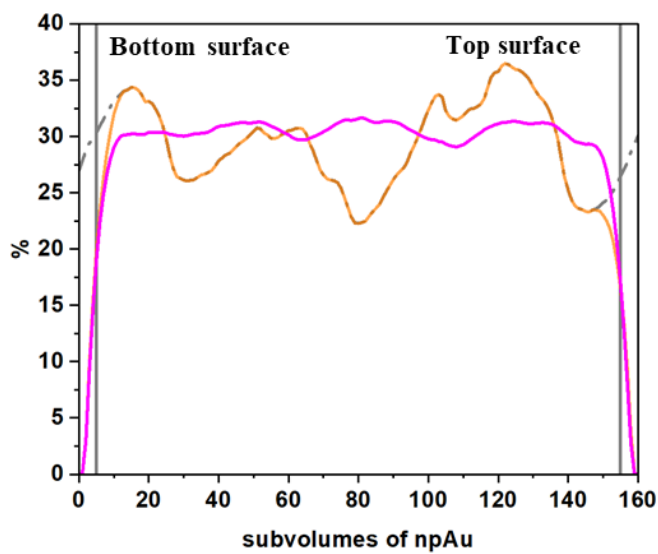

**Figure S3.** The Au volume fraction as a function of depth in the sample (dashed gray curve indicate volume fraction without window function and solid orange curve shows volume fraction after applying window function) the magenta curve shows volume fraction averaged over 15 samples after applying window function. The vertical gray lines indicate the final thickness ~150 nm calculated from half volume fraction ~14 %.

## Supplementary Note 4: Simulated absorption spectra of npAu samples

The simulated absorption spectra of 15 unique npAu samples with the same shell geometry in reciprocal space (see Figure S4). Each of the sample shows distinct resonances in the visible and near-infrared wavelength range. The frequency domain (FD) solver of commercially available CST Studio Suite 2025 was used.[5]

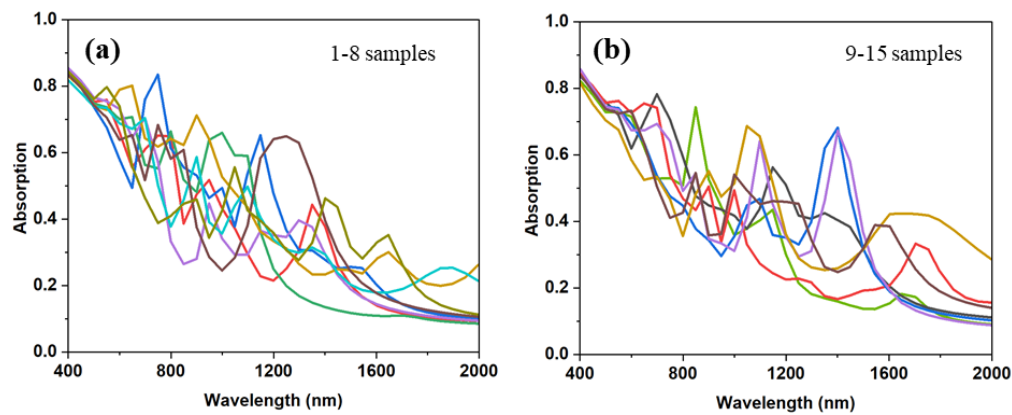

**Figure S4.** The simulated absorption spectra of 15 samples (a) 1-8 samples (b) 9-15 samples show pronounced resonances in visible and near infrared wavelength ranges

## Supplementary Note 5: Resonant and off-resonant electric field distribution of npAu approximants

In Figure S5(a-c), the absorption spectra of three unique realizations with $L_s = 200$ nm and thickness $h = 150$ nm are shown. Each of the samples shows distinct resonances in the visible and NIR wavelength ranges. We analyze the electric field distribution of these samples at resonance and non-resonance wavelengths. Spatially localized electric field is observed at the tips of the two dangling ligaments opposite to each other in all three cases, as shown in Figure S6(a-c). We also compare electric field distributions at non-resonant close-by wavelength where spatially electric field enhancement is absent Figure S6(d-f), confirming that the field enhancements correspond to resonances.

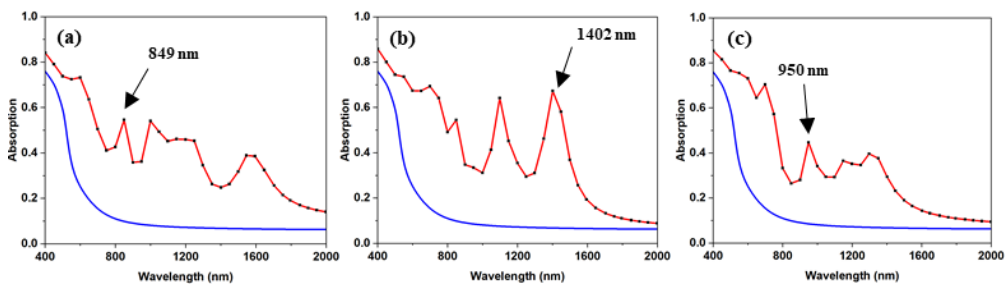

**Figure S5.** (a-c) Three unique npAu approximants show distinct absorption spectra (red with black dots indicating the simulated discrete wavelengths points) across the visible and NIR wavelength range with clearly identifiable resonances. Each sample's absorption spectra are compared to the absorption spectra predicted by the Maxwell-Garnett effective medium model (blue), which serves as a reference.

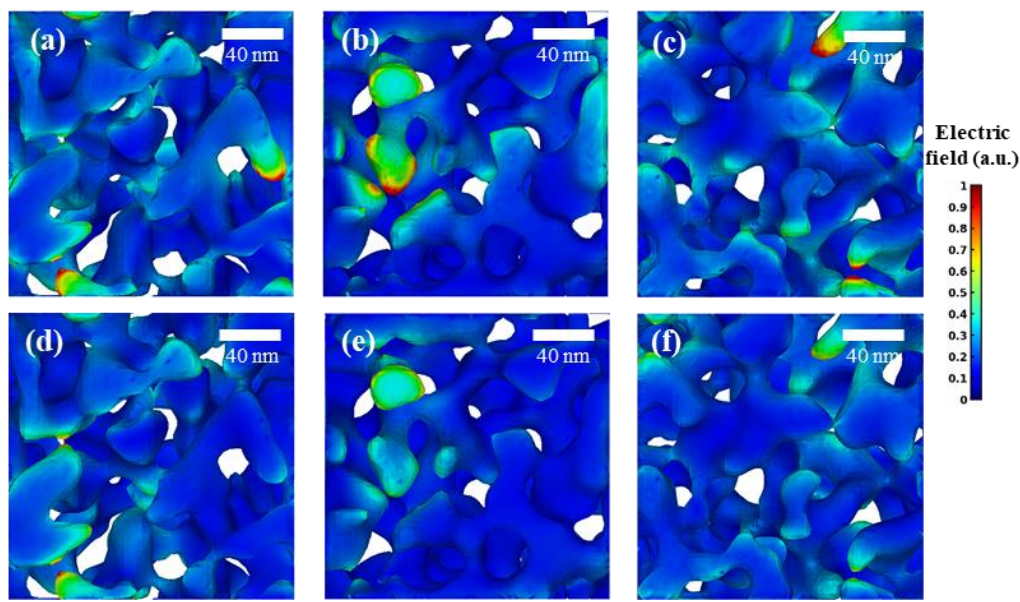

**Figure S6(a-f).** Simulated electric field distributions for three distinct npAu realizations corresponding to absorption peaks at resonant wavelength (a) 849 nm (b) 1402 nm (c) 950 and absorption dips at their corresponding non-resonant wavelength (d) 924 nm (e) 1250 nm (f) 900 nm, respectively. The amplitude of electric field in the metal at the air-gold interface is shown, which helps to illustrate field enhancement at resonant wavelengths in a 3D structure. It is absent when observed at non-resonant wavelengths. NpAu samples are shown at resonant and non-resonant wavelengths for the samples shown in Figure. S5(a-c).

## Supplementary Note 6: Absorption spectra of cylindrical wires grid with a gap

To approach the understanding of observed LSPRs at the surface of npAu, we have simulated a rectangular grid 200 nm by 100 nm of gold cylinders with diameter 15 nm and a varying gap in two configurations. The frequency domain solver was used of commercially available CST

Studio Suite 2025. Plane wave excitation along $z$ direction was used with electric field oriented in $x$ direction along the cylinder with a gap (see Figure S7). Periodic boundary conditions were implemented on lateral boundaries in $x$ and $y$ directions. The gap was introduced in the middle of the wire and at the wire junction. The edges of the cylinders were rounded by a sphere. The simulated structure is free standing in air with refractive index $n = 1$. As the gap decreases, the resonances redshift, which can be explained by the increased capacitance of the gap.[6–8]

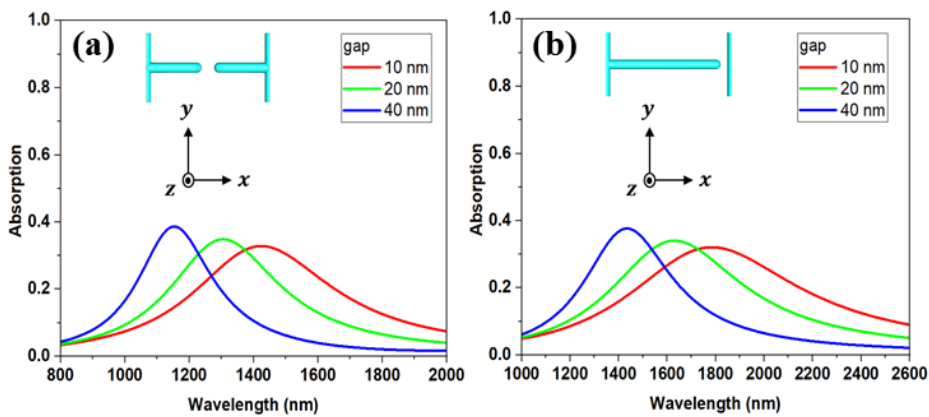

**Figure S7.** The simulated absorption spectra for two configurations of cylinders with different gap sizes show absorption resonance at near infrared wavelength range: (a) gap in the center and (b) gap at the junction. The inserts show the corresponding geometry.

## Supplementary Note 7: Experimental results obtained by UV-Vis spectrometry

The **npAu film optical properties were investigated using UV-Vis spectroscopy while the sample was on a quartz substrate.** In the UV–Vis–NIR band (400 nm - 2000 nm), the substrate is fused silica (amorphous SiO₂), which is essentially lossless and only weakly dispersive. Its primary optical effect is a spectrally flat Fresnel reflection at the air–substrate interfaces, with per-surface reflectance of $\sim 3.3 - 3.5\%$ for $n_{\text{silica}} \sim 1.44 - 1.47$.

Multiple internal reflections within the 1 mm slab form an etalon with free-spectral range well below the spectrometer resolution and therefore incoherently averaged out. Hence, these substrate contributions are broadband, nearly wavelength-independent and averaged by the instrument, the measured spectra predominantly reflect the film's intrinsic response. The substrate mainly introduces an approximately constant transmission offset and a small redistribution between reflection and transmission, leaving absorption effectively unchanged within experimental uncertainty. Therefore, the spectra of npAu film on quartz substrate were compared with simulations of free-standing npAu approximants.

**Due to the observed lateral inhomogeneity within the sample, the diffuse transmission and reflection spectra were measured at three distinct positions using the integrating sphere of the UV-Vis spectrometer. Figure S8 shows the transmission, reflection and absorption spectra of a npAu film measured at three different spots, each of which yielding a slightly different spectrum due to local variation of the thickness and/or filling fraction. Thus, we decided to average the transmission and reflection spectra acquired from three different spots.**

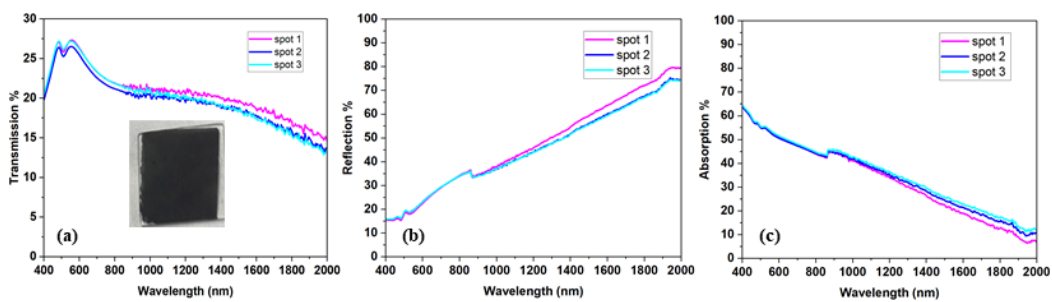

**Figure S8**. The npAu film (a) transmission, (b) reflection and (c) absorption spectra for three different spots of an identical sample. The inset of (a) displays the npAu film. The wavelength ranges from 400 nm to 2000 nm. The step in spectra and increasing noise from 800 nm is due to the detector change.

## Supplementary Note 8: NpAu surface geometry comparison

Figure S9 compares SEM of npAu with a numerically generated npAu approximant. To suppress abrupt truncation at the film boundaries, a window function is applied at the top and bottom surfaces of the approximants (Supporting Information Note 3). The SEM image (Figure S9(a)) reveals highly convex shaped terminations of dangling ligaments, whereas the approximant (Figure S9(b)) exhibits less rounded tips with larger minimum radii, i.e., suppressed curvature extremes. However, a detailed experimental study is required with a focus on npAu surface geometry to emulate it in a numerical model. The systematic disparity in near-surface geometry is expected to affect the optical response, including the density and spectral position of surface-localized resonances.

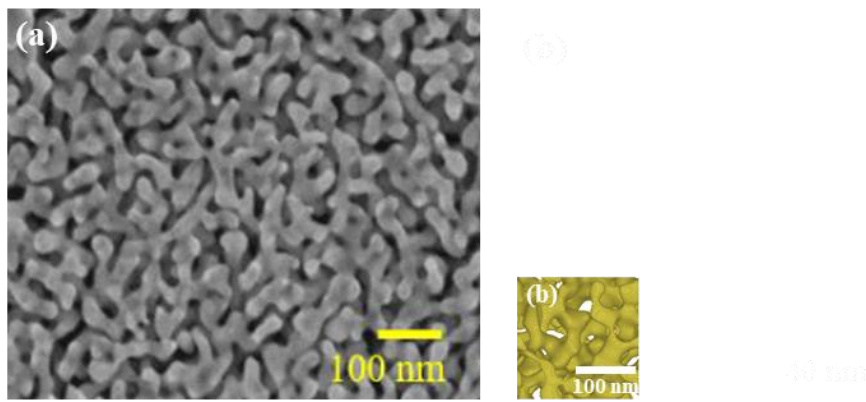

**Figure S9**. (a) SEM of npAu film (b) numerically generated npAu approximant.